\pdfoutput=1
\documentclass[acmsmall,screen,nonacm]{acmart}
\setcopyright{none}
\makeatletter
\@ifundefined{hyxmp@parse@acmart}{}{\let\hyxmp@parse@acmart\relax}
\makeatother
\usepackage[utf8]{inputenc}
\usepackage{microtype}
\usepackage{amsmath}
\usepackage{booktabs,array,tabularx}
\usepackage{pgfplots}
\pgfplotsset{compat=1.18}
\usetikzlibrary{arrows.meta,positioning}
\usepackage{xurl}
\hypersetup{
  hidelinks,
  keeppdfinfo,
  pdftitle={Remote MCP Endpoints After the Stateless Revision: Compatibility and Protocol-Era Transitions},
  pdfauthor={Ziyuan Wang, Genliang Zhu, Xiaobo Chen, Chu Wang},
  pdfkeywords={Model Context Protocol, MCP, stateless protocol, remote endpoints, protocol-era transitions, compatibility, longitudinal measurement}
}
\newcommand{\code}[1]{\texttt{#1}}
\newcommand{\CI}[2]{[#1,\,#2]}

\title[Remote MCP Endpoints After the Stateless Revision]{Remote MCP Endpoints After the Stateless Revision: Compatibility and Protocol-Era Transitions}
\author{Ziyuan Wang}
\affiliation[obeypunctuation=true]{\institution{Independent Researcher}\country{\mbox{}}}
\email{ziyuan.wang.engineer@gmail.com}
\author{Genliang Zhu}
\affiliation{\institution{Accentrust}\city{Vancouver}\state{British Columbia}\country{Canada}}
\affiliation{\institution{Georgia Institute of Technology}\city{Atlanta}\state{Georgia}\country{USA}}
\email{sebby@gatech.edu}
\author{Xiaobo Chen}
\affiliation{\institution{University of Massachusetts Amherst}\city{Amherst}\state{Massachusetts}\country{USA}}
\email{xchen@umass.edu}
\author{Chu Wang}
\affiliation{\institution{Accentrust}\city{Vancouver}\state{British Columbia}\country{Canada}}
\affiliation{\institution{University of Illinois Urbana-Champaign}\city{Urbana}\state{Illinois}\country{USA}}
\email{wallywang@ieee.org}

\authorsaddresses{Authors' Contact Information: Ziyuan Wang, Independent Researcher, ziyuan.wang.engineer@gmail.com; Genliang Zhu, Accentrust, Vancouver, British Columbia, Canada and Georgia Institute of Technology, Atlanta, Georgia, USA, sebby@gatech.edu; Xiaobo Chen, University of Massachusetts Amherst, Amherst, Massachusetts, USA, xchen@umass.edu; Chu Wang, Accentrust, Vancouver, British Columbia, Canada and University of Illinois Urbana-Champaign, Urbana, Illinois, USA, wallywang@ieee.org.}
\date{}

\begin{document}

\begin{abstract}
We measure protocol-era behavior among registry-listed remote Model Context Protocol (MCP) endpoints after the 2026-07-28 stateless revision removed the initialization handshake. We probe a fixed random panel of 1,000 eligible Official MCP Registry entries on 2026-10-24 and 2026-11-24 with one modern \code{server/discover} request and one legacy \code{initialize} request, without credentials or tool invocation. We classify the evidence exposed by each selected endpoint as legacy-observed-only, dual-era, modern-observed-only, or era-unobservable. Among era-observable entries, the legacy-observed-only share is 0.797 in October and 0.729 in November; a domain-cluster sensitivity interval for the November entry-weighted share is [0.592, 0.824]. Of 422 entries that are legacy-observed-only in October and era-observable at both dates, 27 show an observed transition to dual-era evidence. This entry-level transition is concentrated: 16 occur under one domain, and 17 begin with an HTTP 401 response to modern discovery. A conservative sensitivity analysis restricted to the 386 eligible pairs with baseline-observable modern discovery yields 10 newly observed modern-interface transitions. The study estimates endpoint-response compatibility behavior, separating protocol-era visibility from release provenance. For clients seeking broad reach across the sampled population, the observed prevalence supports retaining legacy initialization fallback. An earlier preprint probed the same registry four weeks after the revision~\cite{sunkavalli2026deploymentgap}; both studies find legacy-observed-only behavior dominant. Data and code are public.
\end{abstract}

\maketitle

\section{Introduction}

The Model Context Protocol (MCP) connects AI applications to external tools and data sources~\cite{anthropic2024mcp}. Its 2026-07-28 revision changes how clients begin communicating with an endpoint: the stateless core removes the initialization handshake and protocol-level session identifier, and requests carry their version and client capabilities as metadata~\cite{mcp2026changelog}. A remote endpoint can continue exposing legacy requests, expose the modern protocol, or expose both eras. The distribution of observable behaviors determines how broadly a client can connect using a particular era.

Early adoption reports describe implementation by infrastructure providers and continuing backward compatibility~\cite{aaif2026ecosystemresponds,contextiq2026mcpexplained}, but do not estimate protocol-era behavior in a random sample of registry entries. Sunkavalli's preliminary preprint does: about four weeks after the revision, it probed a random panel of the Official MCP Registry with the same discovery and initialization pair, plus a same-session negative control, and found that 3 of 188 probeable servers in a 400-server panel spoke the new protocol~\cite{sunkavalli2026deploymentgap}.

We measure that distribution later in the revision's deployment period and its subsequent change. The study follows a fixed panel of 1,000 entries in the Official MCP Registry, each represented by a selected remote Streamable HTTP endpoint, from October to November. The primary unit is a registry-entry/selected-endpoint observation. This unit estimates externally visible protocol behavior; release provenance, operator identity, software version, and SDK provenance form separate measurement targets. Our scope is behavior visible without authentication at registry-listed remote endpoints; authenticated services and source-only distributions are outside the era estimates.

We probe the same 1,000 selected endpoints twice, one month apart, and report cross-sectional distributions, the full transition matrix, and changes in the era-observable population. A post hoc analysis describes concentration by domain and shows how clustering changes uncertainty. Relative to the first wave of Sunkavalli's panel, we add observations three and four months after the revision and paired transitions of the same entries one month apart. Like that study, we find legacy-observed-only behavior dominant where an era is observable. Throughout, an \emph{observed protocol-era transition} is the entry/endpoint response transition from legacy-observed-only evidence to dual-era or modern-observed-only evidence. It is the paper's compatibility outcome; software, SDK, deployment, and operator attribution require a separate provenance design.

This paper makes three contributions. First, it provides a fixed-panel measurement design that preserves every sampled registry entry across two waves, including authentication and reachability outcomes. Second, it formalizes protocol-era visibility and reports the complete six-state transition matrix, separating paired endpoint-response changes from changes in observability. Third, it quantifies the concentration and authentication sensitivities of the observed transition signal, translating the measurement into a conservative fallback recommendation for clients that must cover the sampled endpoint population.

\section{Related Work}

Existing MCP measurements provide useful sampling and probing precedents. Hasan et al. study the security and maintainability of open-source MCP servers~\cite{hasan2026modelcontextprotocolmcp}. Bharti follows the Official MCP Registry longitudinally to measure changes in descriptions and descriptor fields~\cite{bharti2026registrydescriptionsstaleunevenly}. That study measures registry metadata rather than live protocol behavior, and its observation period ends on the stateless revision's release date. Zhou et al. identify remote MCP servers through active initialization probes and study authentication security~\cite{zhou2026measurementstudyauthenticationsecurity}. Their legacy initialization criterion predates the revision. Sohail characterizes hosting centralization and authentication among remote endpoints listed in the official registry and Smithery, also before the revision~\cite{sohail2026characterizingnetworkcentralizationobservability}. A discovery probe is also needed to observe servers that expose only the modern interface. Afsar draws a seeded random sample of npm-distributed local servers from the same registry, probes each draw once, and records an outcome for every draw~\cite{afsar2026randomdrawmcpregistry}; we apply the same accounting to remote endpoints.

MCP-Eco links servers, consumers, and development activity, providing broader ecosystem context~\cite{tounoussi2026mcpecolinkeddatasetmcp}. Our outcome is the protocol era exposed by the same endpoint at successive observations. The closest prior measurement of that outcome is Sunkavalli's preprint~\cite{sunkavalli2026deploymentgap}. It samples the same registry with the same probe pair plus a same-session negative control, records unreachable and authentication-walled servers separately, and treats publishers as the unit of adoption. Its public results cover the first corrected wave of a planned multi-wave panel, about four weeks after the revision. The adoption and endpoint-probing reports above discuss the outcome without random sampling~\cite{aaif2026ecosystemresponds,contextiq2026mcpexplained}. We add observations three and four months after the revision and paired month-over-month transitions of the same entries, and we report domain-cluster bootstrap intervals for entry-weighted shares. Both studies find legacy-observed-only behavior dominant where an era is observable. Their shares are not directly comparable because dates, sampling frames, denominators, and controls differ; of the 991 panel entries with a retrievable registry history, 581 were first published after that study's first wave.

Our design follows longitudinal Internet measurements of protocol transitions, which observe the adoption of a new version repeatedly over time: IPv6~\cite{Czyz_2014}, TLS~\cite{Kotzias_2018}, early TLS~1.3, where a few large operators drove much of the observed uptake~\cite{Holz_2020}, and QUIC, where large operators changed configurations across their deployments almost at once while self-managed deployments upgraded over years~\cite{Zirngibl_2021,Kempf_2026}. Wickramasinghe et al.\ reuse an identical domain set across three scans of post-quantum TLS deployment~\cite{Wickramasinghe_2026}. Our panel likewise fixes the sampled entries, but it keeps every entry and reports those that become era-unobservable instead of restricting analysis to entries observed successfully at every wave. Within MCP, Guo et al.\ found that only 95 of 341 clients supported the newer Streamable HTTP transport in August 2025 and concluded that clients had not kept pace with protocol evolution~\cite{guo2025measurementstudymodelcontext}.

Software-engineering studies report slow downstream adoption of breaking changes while the old interface keeps working. Sawant et al.\ report that most clients of popular Java APIs and the JDK freeze to one API version~\cite{Sawant_2017}. About eight years after Python~3, 49 of 51 studied applications remained compatible with Python~2, and only two had dropped it~\cite{Malloy_2017}. Even when hosted LLM models are retired and old requests fail, most observed downstream migrations occur after the retirement date~\cite{kim2026modelretiresempiricalstudy}. These studies concern source code and mostly lagging clients, whereas we observe responses from deployed endpoints; we cite them as context, not as predictions for MCP. The MCP versioning specification defines dual-era servers and sets no date for retiring legacy versions~\cite{mcp2026versioning}.

\section{Study Design}
\label{sec:design}

\subsection{Sampling and repeated observations}

We use the Official MCP Registry's latest-version snapshot collected on 2026-10-24~\cite{mcpregistry}. The snapshot contains 49,562 records. We retain entries marked latest and active with an eligible remote Streamable HTTP endpoint~\cite{mcp2026streamablehttp}: an HTTPS URL without template variables. This yields 31,326 candidates. Each entry contributes its first eligible URL.

A feasibility sample of 50 entries, drawn after sorting names with random seed 2303, yielded 36 era-observable entries (0.72). We exclude that pilot from the main panel, leaving 31,276 candidates, and draw 1,000 entries using seed 2304. The pilot establishes feasibility and does not contribute to the main estimates.

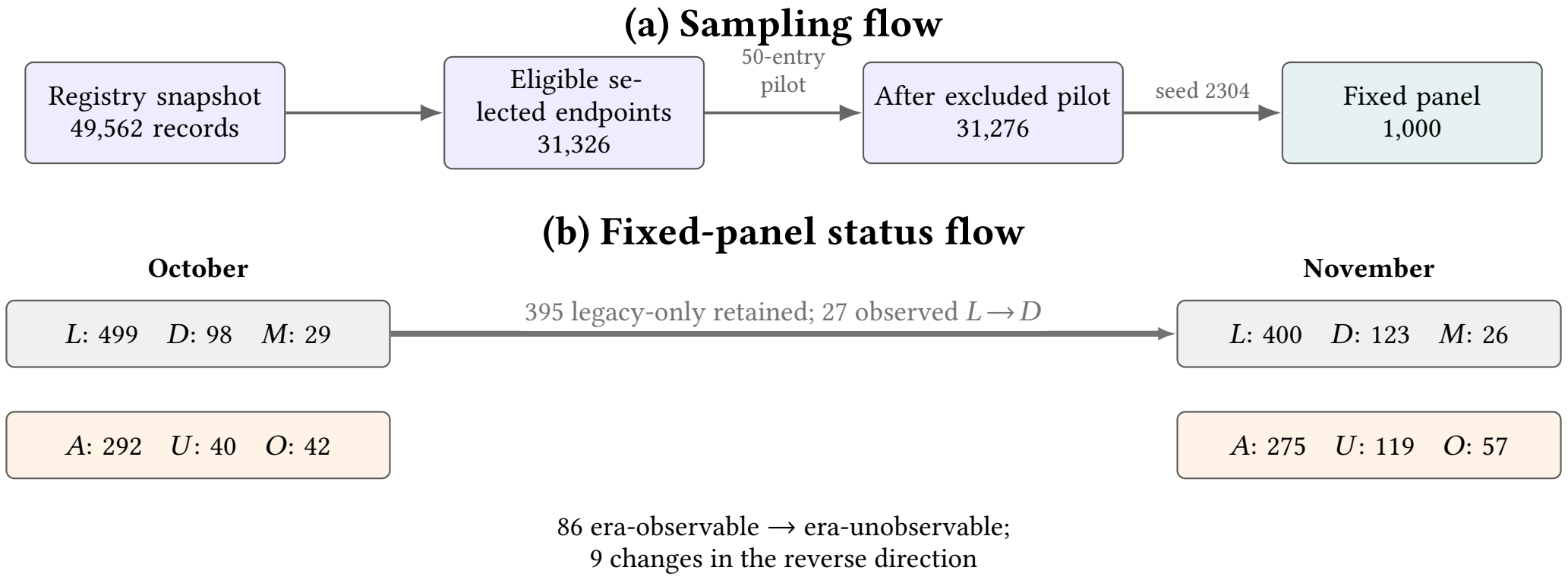
\begin{figure*}[t]
\centering
\textbf{(a) Sampling flow}\\[-0.35em]
\begin{tikzpicture}[font=\scriptsize,>=Latex,
  flowbox/.style={draw=black!65,rounded corners=2pt,fill=blue!7,align=center,
    text width=2.05cm,minimum height=0.88cm,inner sep=3pt},
  arrow/.style={-{Latex[length=2mm]},line width=0.7pt,black!60}]
\node[flowbox] (r) at (0,0) {Registry snapshot\\49,562 records};
\node[flowbox] (e) at (3.65,0) {Eligible selected endpoints\\31,326};
\node[flowbox] (p) at (7.30,0) {After excluded pilot\\31,276};
\node[flowbox,fill=teal!10] (s) at (10.95,0) {Fixed panel\\1,000};
\draw[arrow] (r)--(e);
\draw[arrow] (e)--(p) node[midway,above,font=\tiny,align=center]{50-entry\\pilot};
\draw[arrow] (p)--(s) node[midway,above,font=\tiny]{seed 2304};
\end{tikzpicture}\\[0.65em]
\textbf{(b) Fixed-panel status flow}\\[-0.35em]
\begin{tikzpicture}[font=\scriptsize,>=Latex,
  status/.style={draw=black!65,rounded corners=2pt,align=center,text width=3.20cm,
    minimum height=0.58cm,inner sep=2pt},
  stay/.style={-{Latex[length=1.6mm]},line width=1.5pt,black!55}]
\node[font=\scriptsize\bfseries] at (0,1.30) {October};
\node[font=\scriptsize\bfseries] at (10.2,1.30) {November};
\node[status,fill=gray!12] (ol) at (0,0.72) {$L$: 499\quad $D$: 98\quad $M$: 29};
\node[status,fill=orange!10] (ou) at (0,-0.25) {$A$: 292\quad $U$: 40\quad $O$: 42};
\node[status,fill=gray!12] (nl) at (10.2,0.72) {$L$: 400\quad $D$: 123\quad $M$: 26};
\node[status,fill=orange!10] (nu) at (10.2,-0.25) {$A$: 275\quad $U$: 119\quad $O$: 57};
\draw[stay] (ol.east) -- node[above,font=\scriptsize,fill=white,inner sep=1pt]{395 legacy-only retained; 27 observed $L\!\rightarrow\!D$} (nl.west);
\node[font=\scriptsize,align=center] at (5.10,-1.12) {86 era-observable $\rightarrow$ era-unobservable;\\9 changes in the reverse direction};
\end{tikzpicture}
\caption{Study flow and major fixed-panel status flows. $L$, $D$, and $M$ denote legacy-observed-only, dual-era, and modern-observed-only evidence; $A$, $U$, and $O$ denote authentication-blocked, unreachable, and other evidence. Panel (b) highlights the flows most relevant to interpretation; Table~\ref{tab:transitions} reports the complete six-state transition matrix.}
\Description{Panel (a) shows the sampling flow from 49,562 registry records to a 1,000-entry fixed panel after filtering and excluding a 50-entry pilot. Panel (b) compares October and November status counts, highlighting 395 retained legacy-only entries, 27 observed transitions from legacy-only to dual-era, 86 entries that become era-unobservable, and 9 in the reverse direction.}
\label{fig:design-flow}
\end{figure*}

The main observations occur on 2026-10-24 and 2026-11-24, approximately 31 days apart and about three and four months after the revision, respectively. The follow-up reuses the same sample, URLs, and probe implementation. Both waves account for all 1,000 sampled entries. A failed request remains an observed outcome; we do not replace the entry. Figure~\ref{fig:design-flow} summarizes the sampling path and the principal status flows retained by this fixed-panel design.

\begin{table}[t]
\centering
\caption{Cross-sectional observed counts using prespecified categories in the fixed panel. October is 2026-10-24; November is 2026-11-24. The first three categories have observable protocol-era evidence.}
\label{tab:counts}
\begin{tabular}{lrr}
\toprule
Observed category & October & November \\
\midrule
Legacy-observed-only & 499 & 400 \\
Dual-era & 98 & 123 \\
Modern-observed-only & 29 & 26 \\
Auth & 292 & 275 \\
Unreachable & 40 & 119 \\
Other & 42 & 57 \\
\midrule
All entries & 1,000 & 1,000 \\
Era-observable entries & 626 & 549 \\
\bottomrule
\end{tabular}
\end{table}

\subsection{Protocol probes and classification}

The protocol distinguishes legacy initialization-based versions, modern versions with metadata on each request, and dual-era implementations~\cite{mcp2026versioning}. We operationalize observable evidence for these categories with paired JSON-RPC requests~\cite{jsonrpc2}. The modern probe sends \code{server/discover} with protocol version \code{2026-07-28}, empty client capabilities, and client identity in request metadata, together with the required version and method headers. A JSON-RPC result object containing a \code{supportedVersions} field supplies modern-era evidence~\cite{mcp2026discover}. The classifier does not validate the listed versions; every such result in both waves listed \code{2026-07-28}. The classifier, fixed before the first wave, also accepts recognized modern protocol errors; none occurred in either wave. Unlike Sunkavalli~\cite{sunkavalli2026deploymentgap}, we sent no same-session negative control, so an endpoint that returned such a field for arbitrary methods would be recorded with modern-era evidence. The paired probe establishes an observable response signature; full conformance and end-to-end client interoperability are not measured.

The legacy probe sends \code{initialize}, requests version \code{2025-11-25}, and supplies empty capabilities and client identity. A result containing \code{protocolVersion} supplies legacy evidence. If initialization returns a session identifier, the probe attempts to close that session. We parse JSON responses and the first data message of an event-stream response. We do not invoke tools, send credentials, or attempt to bypass authentication.

An entry is \emph{legacy-observed-only} when only initialization supplies era evidence, \emph{modern-observed-only} when only discovery does, and \emph{dual-era} when both do. These categories comprise the \emph{era-observable} set; modern-observed-only and dual-era together have \emph{modern-era evidence}. When neither probe supplies era evidence, we assign \emph{auth} if either returns HTTP 401 or 403, \emph{unreachable} if both encounter connection failures, timeouts, or server errors, and \emph{other} otherwise. Successful era evidence takes precedence over a failure on the other interface.

This precedence matters for interpretation. In October, 37 legacy-observed-only entries returned HTTP 401 to discovery while initialization succeeded. In November, 23 returned 401 and one returned 403. Such classifications describe the unauthenticated interface that we observed; the modern interface may behave differently with credentials.

For entry $i$ and observation date $t$, let
$Y_{it}\in\{L,D,M,A,U,O\}$ denote, respectively, legacy-observed-only,
dual-era, modern-observed-only, authentication-blocked, unreachable, and other
evidence. Let $C_{it}=\mathbb{1}\{Y_{it}\in\{L,D,M\}\}$ indicate whether an
era is observable under the unauthenticated paired probes. The conditional
legacy-observed-only prevalence is
\[
\widehat p_L(t)=
\frac{\sum_i \mathbb{1}\{Y_{it}=L\}}{\sum_i C_{it}}.
\]
We define
\[
T_i=\mathbb{1}\{Y_{i,\mathrm{Oct}}=L,\;
Y_{i,\mathrm{Nov}}\in\{D,M\}\}
\]
as an \emph{observed protocol-era transition}. $T_i$ is an endpoint-response
outcome; software, SDK, deployment, and operator provenance require separate
release-level evidence.

\begin{table*}[t]
\centering
\caption{Main prevalence estimates, paired change, and observed protocol-era transition. All intervals are 95\%. Entry-level intervals use Wilson for proportions and bootstrap for the paired change. Cross-sectional metrics were fixed before the October wave; paired analyses were fixed after that wave and before November results were inspected. Domain-cluster intervals were added after inspection of the November collection and are descriptive. Both methods estimate entry-weighted proportions. The transition denominator is the 422 initially legacy-observed-only entries with observable eras at both dates.}
\label{tab:intervals}
\small
\setlength{\tabcolsep}{3pt}
\begin{tabularx}{\textwidth}{@{}>{\raggedright\arraybackslash}X >{\raggedright\arraybackslash}p{2.55cm} r l l@{}}
\toprule
Outcome and denominator & Observation & Estimate & Entry-level interval & Domain-cluster interval \\
\midrule
Legacy-observed-only / era-observable & October & 0.797 & \CI{0.764}{0.827} & \CI{0.685}{0.870} \\
 & November & 0.729 & \CI{0.690}{0.764} & \CI{0.592}{0.824} \\
Modern-era evidence / 1,000 & October & 0.127 & \CI{0.108}{0.149} & \CI{0.085}{0.187} \\
 & November & 0.149 & \CI{0.128}{0.172} & \CI{0.099}{0.216} \\
Paired modern-era-evidence change / 1,000 & November minus October & $+0.022$ & $[+0.010,\,+0.034]$ & $[-0.001,\,+0.061]$ \\
Observed $L\!\rightarrow\!\{D,M\}$ transition / 422 & October to November & 0.064 & \CI{0.044}{0.091} & \CI{0.016}{0.164} \\
\bottomrule
\end{tabularx}
\end{table*}

\begin{table*}[t]
\centering
\caption{Paired era transition matrix. Rows are observations on 2026-10-24 and columns on 2026-11-24. An observed protocol-era transition is restricted to legacy-observed-only $\rightarrow$ dual-era or modern-observed-only. Auth, unreachable, and other are era-unobservable outcomes.}
\label{tab:transitions}
\small
\setlength{\tabcolsep}{3.5pt}
\begin{tabular}{lrrrrrrr}
\toprule
October $\backslash$ November & Legacy & Dual & Modern & Auth & Unreach. & Other & Total \\
\midrule
Legacy-only & 395 & 27 & 0 & 6 & 62 & 9 & 499 \\
Dual-era & 0 & 91 & 0 & 1 & 6 & 0 & 98 \\
Modern-only & 0 & 1 & 26 & 1 & 0 & 1 & 29 \\
Auth & 4 & 2 & 0 & 266 & 14 & 6 & 292 \\
Unreachable & 1 & 2 & 0 & 0 & 34 & 3 & 40 \\
Other & 0 & 0 & 0 & 1 & 3 & 38 & 42 \\
\midrule
Total & 400 & 123 & 26 & 275 & 119 & 57 & 1,000 \\
\bottomrule
\end{tabular}
\end{table*}

\subsection{Outcomes and analysis timing}

The era definitions and cross-sectional metrics were fixed before the October wave. Our primary outcome is the legacy-observed-only share among era-observable entries. We prespecified a threshold of at least 0.30 for a substantial compatibility gap and a stopping threshold below 0.10. This endpoint-level quantity operationalizes protocol-era coverage. The specification defines the compatibility relationship between modern-only clients and legacy-only servers~\cite{mcp2026versioning}; product-level interoperability is a distinct end-to-end validation target.

We retain the original entry-level Wilson 95\% intervals~\cite{Wilson_1927}. The paired analyses were defined after seeing October data and before any November record was inspected: the transition script was hashed at 11:48:42 UTC on November 24, after the second wave started at 11:45:32 and before it finished at 11:51:38. These analyses comprise the full era transition matrix, modern-era-evidence prevalence over the fixed panel and its paired change, observed transitions among entries era-observable at both waves, and entry into or departure from the era-observable set. For the paired prevalence change, the plan specifies an entry-level bootstrap with 10,000 resamples~\cite{Efron_1979} and an exact McNemar test~\cite{McNemar_1947}.

Because the revision moves client capabilities from the handshake into each request, an endpoint can return different tools depending on the declared capabilities. The capability-declaration check sends \code{tools/list} to entries with modern-era evidence with empty \code{clientCapabilities} and with \code{elicitation} declared as an empty object. When both calls succeed, it compares returned tool names while ignoring order.

After inspecting November results, we added analyses of domain clustering, hostname deduplication, an outage-sensitive subset, and the distribution of changes across domains. On November 25, we added the domain-cluster interval for the observed-transition share and the sensitivity analysis excluding the largest changing domain. On November 29, we added a check of the endpoints listed in registry version histories and a breakdown of transitions by their October discovery status. All of these analyses are post hoc and descriptive. We present domain-cluster intervals alongside the original entry-level intervals because the panel contains entries sharing infrastructure. The original outcomes and thresholds remain unchanged.

\paragraph{Measurement ethics.}
Each wave used eight workers across entries; requests within an entry were serial, with one-second pauses between requests and a 20-second timeout for POST requests. For each entry, the probe attempted one discovery and one initialization request, plus two tool-list requests for endpoints with modern-era evidence; an issued session identifier triggered a DELETE cleanup request with a ten-second timeout. These requests inspected protocol metadata and tool listings without invoking tools, sending credentials, or attempting authentication bypass; initialization could create a temporary session. Following measurement-ethics guidance~\cite{Partridge_2016}, interaction was limited to protocol metadata and tool listings. The User-Agent string identified the traffic as research measurement but contained no contact address, repository URL, or opt-out mechanism, which limited operators' ability to contact us or opt out.

\section{Main Results}

\subsection{Legacy-observed-only behavior remains common}

\begin{figure}[t]
\centering
\begin{tikzpicture}
\begin{axis}[
  width=0.75\columnwidth,height=4.8cm,
  xmin=0,xmax=1,ymin=0.5,ymax=3.5,
  xlabel={Entry-weighted share},
  ytick={1,2,3},yticklabels={Observed transition,Legacy-observed-only Nov.,Legacy-observed-only Oct.},
  xtick={0,0.2,0.4,0.6,0.8,1},
  tick label style={font=\footnotesize},label style={font=\footnotesize},
  axis y line*=left,axis x line=bottom,
  y axis line style={draw=none},ytick style={draw=none},
  legend style={at={(0.5,-0.42)},anchor=north,draw=none,font=\footnotesize},
  legend cell align=left,
]
\addplot+[only marks,mark=*,mark size=1.6pt,black,
  error bars/.cd,x dir=both,x explicit,error bar style={line width=0.8pt}]
  coordinates {(0.064,1.1) += (0.027,0) -= (0.020,0)
               (0.729,2.1) += (0.035,0) -= (0.039,0)
               (0.797,3.1) += (0.030,0) -= (0.033,0)};
\addlegendentry{Entry-level Wilson 95\%}
\addplot+[only marks,mark=square*,mark size=1.6pt,blue!65!black,
  error bars/.cd,x dir=both,x explicit,error bar style={line width=0.8pt}]
  coordinates {(0.064,0.9) += (0.100,0) -= (0.048,0)
               (0.729,1.9) += (0.095,0) -= (0.137,0)
               (0.797,2.9) += (0.073,0) -= (0.112,0)};
\addlegendentry{Domain-cluster 95\% (post hoc)}
\end{axis}
\end{tikzpicture}
\caption{Uncertainty for legacy-observed-only shares among era-observable entries and observed $L\!\rightarrow\!\{D,M\}$ transitions among 422 initially legacy-observed-only entries with observable eras at both dates. Points show the same estimates for both methods; vertical offsets separate their 95\% intervals. Domain-cluster intervals are descriptive.}
\Description{A horizontal interval plot compares entry-level Wilson intervals and post hoc domain-cluster intervals for legacy-observed-only shares in October and November and for the observed transition share. The domain-cluster intervals are visibly wider for all three estimates.}
\label{fig:intervals}
\end{figure}
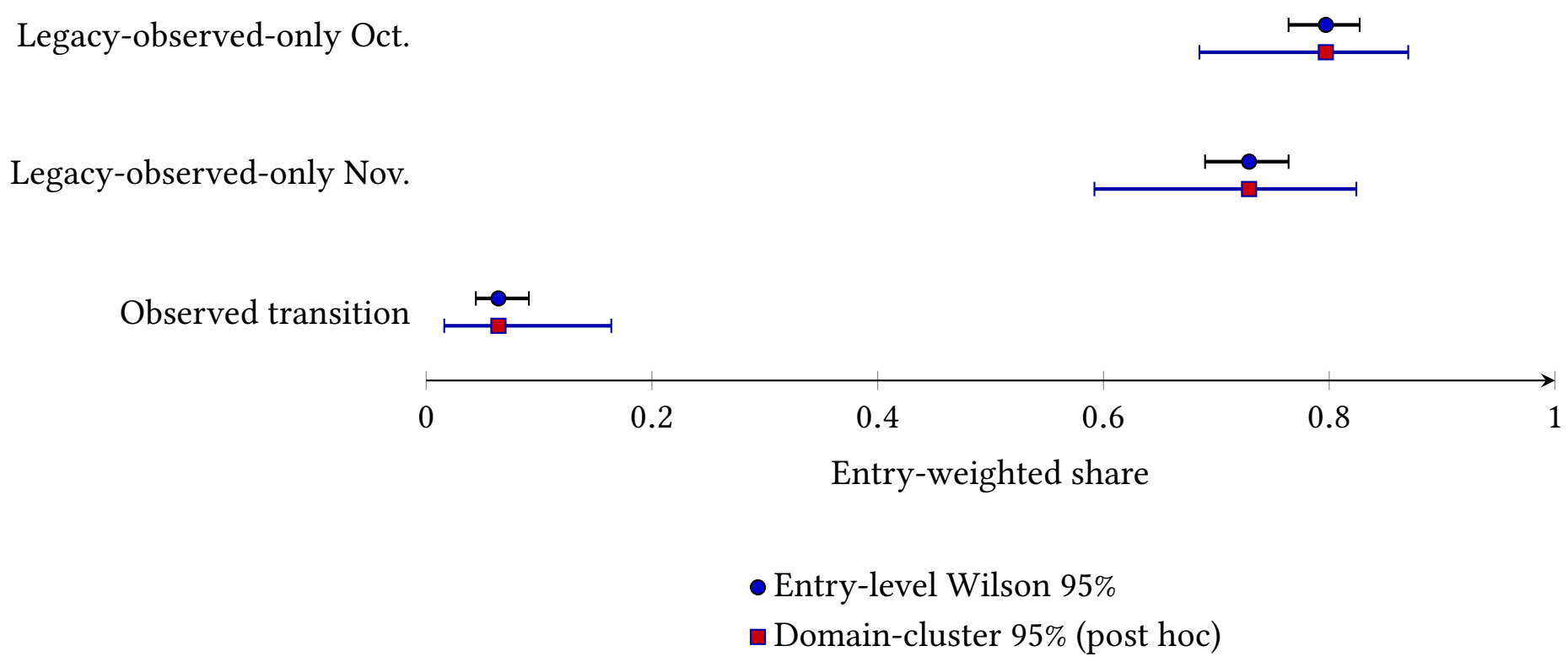

Table~\ref{tab:counts} gives the complete observed distributions. The era-observable share falls from 0.626 to 0.549. Among those entries, the legacy-observed-only share is 0.797 in October and 0.729 in November (Table~\ref{tab:intervals}). The domain-cluster intervals (post hoc) are \CI{0.685}{0.870} and \CI{0.592}{0.824}, respectively. Both lower bounds exceed the prespecified 0.30 threshold. Figure~\ref{fig:intervals} compares the entry-level and domain-cluster intervals.

The complementary modern-era-evidence share among era-observable entries rises from 0.203, with entry-level interval \CI{0.173}{0.236}, to 0.271, \CI{0.236}{0.310}. These conditional shares describe different era-observable sets at the two dates. The fixed-panel and paired results below distinguish observed endpoint-response changes from changes in visibility.

Legacy version negotiation also persists within the initialization interface. In November, successful initialization responses report \code{2025-11-25} for 338 entries, \code{2025-03-26} for 103, \code{2025-06-18} for 65, and \code{2024-11-05} for 17. Among successful discovery responses, 112 advertise only \code{2026-07-28}, while 37 also list older versions.

\subsection{Paired transitions and visibility}

\paragraph{Full panel.}
Table~\ref{tab:transitions} retains every sampled entry. Modern-era-evidence counts increase from 127 to 149. There are 31 entries newly observed with modern-era evidence and nine no longer observed with modern-era evidence. The paired increase over the fixed denominator is $+0.022$. Its entry-level bootstrap interval is $[+0.010,+0.034]$, and the exact McNemar test gives $p=0.00068$. The domain-cluster interval (post hoc) is wider, $[-0.001,+0.061]$, and includes zero. The McNemar result treats entries as independent pairs and does not adjust for common operators.

\paragraph{Observed protocol-era transitions.}
There are 540 entries era-observable at both dates. Within this set, the legacy-observed-only share falls from 0.781, with entry-level Wilson interval \CI{0.745}{0.814}, to 0.731, \CI{0.693}{0.767}. Of the 422 initially legacy-observed-only entries in this set, 27 become dual-era and none become modern-observed-only. The observed-transition share is 0.064, with entry-level Wilson interval \CI{0.044}{0.091} and domain-cluster interval \CI{0.016}{0.164} (post hoc). Sixteen of the 27 transitions occur under the largest changing domain; excluding that domain leaves 11 transitions among 406 eligible entries, a share of 0.027. No entry with modern-era evidence becomes legacy-observed-only among the era-observable pairs. The transition denominator excludes initially legacy-observed-only entries that become era-unobservable; their subsequent eras are unobserved. Section~\ref{sec:posthoc} describes the October discovery status of the 27 transitions.

Table~\ref{tab:transition-sensitivity} makes the evidence hierarchy explicit. The 27/422 result remains the primary observed endpoint-response transition. Of those 422 eligible pairs, 36 had an HTTP 401 or 403 response to modern discovery at baseline, leaving 386 pairs with baseline-observable modern discovery; the corresponding conservative sensitivity is 10/386 newly observed modern-interface transitions. The rows characterize response visibility and compatibility behavior; software or SDK release attribution requires separate provenance evidence.

\begin{table}[t]
\centering
\caption{Observed-transition sensitivities. All rows describe registry-entry/selected-endpoint observations. The latter three rows are post hoc descriptive analyses; release-level attribution is outside their estimand.}
\label{tab:transition-sensitivity}
\small
\begin{tabular}{p{0.56\columnwidth}rr}
\toprule
Definition & Count & Share \\
\midrule
Observed $L\!\rightarrow\!\{D,M\}$ among era-observable pairs & 27/422 & 0.064 \\
Baseline modern discovery observable (excluding 36 blocked pairs) & 10/386 & 0.026 \\
Excluding largest changing domain & 11/406 & 0.027 \\
Initially $L$ and later era-unobservable & 77/499 & 0.154 \\
\bottomrule
\end{tabular}
\end{table}

\paragraph{Changes in observability.}
Of the October era-observable entries, 86 become era-unobservable: 68 are unreachable, ten are other, and eight require authentication. In the opposite direction, nine entries become era-observable: three previously unreachable and six previously in auth. The cross-sectional change therefore combines endpoint-response behavior with changes in visibility, so we report it separately from the paired transition estimand. Likewise, the 31 newly modern-era-evidence observations include entries whose earlier era was unknown; only the initially legacy-observed-only transitions enter the primary observed-transition definition.

Among entries with successful initialization at both dates, 491 retain the same negotiated version. Nineteen change from \code{2025-06-18} to \code{2025-11-25}, and one from \code{2024-11-05} to \code{2025-11-25}. The two October responses that reported \code{2026-07-28} through initialization report \code{2025-11-25} in November. All 19 entries that changed from \code{2025-06-18} are among the 27 observed transitions (Section~\ref{sec:posthoc}); the version change itself does not enter the transition count, which requires new modern-interface evidence.

\subsection{Selected tool-name comparison under two client declarations}

Both tool-list requests succeed on 95 of the 127 October endpoints with modern-era evidence and on 108 of the 149 corresponding November endpoints. Under the two selected client declarations, returned tool names agree on every successful paired response: 0/95 in October and 0/108 in November. The corresponding one-sided 95\% binomial upper bounds for an observed tool-name difference are 3.1\% and 2.7\%. This diagnostic isolates returned tool names; schema behavior and executable functionality are separate protocol properties. The counts cover successful paired responses, with entries sharing operators.

\section{Post Hoc Descriptive Analyses}
\label{sec:posthoc}

The November results revealed shared infrastructure behind many registry entries. We added the following descriptive analyses after inspecting those results to assess dependence and sensitivity.

\paragraph{Domain-cluster uncertainty.}
We group entries by the last two labels of the URL hostname, producing 685 clusters, then resample whole clusters with replacement for 10,000 bootstrap replicates using seed 2303~\cite{Field_2007}. Each replicate retains all entries and both observations within each selected cluster, and recomputes the entry-weighted proportions. For observed transitions, we resample the same 685 clusters and recompute the numerator and eligible-pair denominator within each replicate, including clusters that contribute no eligible pairs. We take the 2.5th and 97.5th percentiles at zero-based sorted indices 249 and 9,749. Table~\ref{tab:intervals} reports the resulting intervals. This suffix-based cluster is a conservative dependence sensitivity rather than an ownership map: it uses neither the Public Suffix List nor registry publisher identities, and it groups all 110 entries on one shared platform suffix together.

\paragraph{One entry per hostname.}
Keeping the first entry in the recorded sample order for each full hostname yields 791 entries. The legacy-observed-only share among era-observable entries is 0.757 in October and 0.683 in November. Newly observed modern-era-evidence and no-longer-observed modern-era-evidence counts remain 31 and nine. Hostname deduplication therefore changes prevalence estimates without removing the concentration of changes: the 16 transitions under the largest changing domain use different subdomains. The public repository includes the sampled-panel order used for this tie-break.

\paragraph{Outage sensitivity.}
One host (\emph{Host B}) accounts for 47 November TLS failures. Of these entries, 46 were legacy-observed-only in October and one was other. A descriptive sensitivity analysis excluding these 47 entries leaves 953 entries; unreachable counts then rise from 40 to 72 rather than from 40 to 119. In this subset, the era-observable counts are 580 and 549, and the legacy-observed-only shares are 0.781 and 0.729. The hostname is omitted because a short unauthenticated availability observation is not an attribution of service quality.

\paragraph{Where changes occur.}
New modern-era-evidence observations span 16 domain clusters. The largest changing domain (\emph{Domain A}) contributes 16 of the 31 entries, and each remaining cluster contributes one. All 16 Domain A entries move from legacy-observed-only to dual-era, accounting for 16 of the 27 observed transitions. The nine entries no longer observed with modern-era evidence span nine domains and become unreachable (six), auth (two), or other (one). Their modern support is unobserved at follow-up. The domain label is withheld because the estimand is entry/endpoint response behavior rather than a named operator release.

\paragraph{October baseline of the observed transitions.}
Seventeen of the 27 transitions, including all 16 under Domain A, come from entries whose discovery request returned HTTP 401 in October while initialization succeeded; the classification rule records them as legacy-observed-only, while their October modern support remains unobserved. Among the 422 eligible pairs, 36 have an HTTP 401 or 403 response to baseline discovery; excluding those pairs leaves 10 transitions among 386 eligible entries (0.026). The retained responses record a changed version value for all 16 Domain A entries in November (22 of the 27 transitions did, compared with 77 of the 395 eligible entries that did not transition), and all 19 entries that changed their negotiated legacy version from \code{2025-06-18} to \code{2025-11-25} are among the 27 transitions. Together these signals locate a rollout-aligned change in endpoint responses. Distinguishing access-control, gateway, deployment, code, and SDK provenance is a separate release-level measurement task.

\paragraph{Listed endpoint changes.}
Registry entries can change or drop their remote endpoint while keeping their name~\cite{kraishan2026namedifferentserversecurity}, and the November wave re-probed the October URLs. On November 29 we retrieved each entry's version history from the registry and took the latest version published before the November wave started. For 984 entries, that version still listed the probed URL; six listed a different URL and one listed no eligible remote endpoint. Nine entries were absent from the registry when checked, so their removal dates are unknown. Of the 86 entries that became era-unobservable, 78 still listed the probed URL, five were absent, and three had moved to new URLs; all three belong to one publisher, were legacy-observed-only in October, and returned authentication errors at their old URLs in November. We did not probe the new URLs. If all three had exposed modern-era evidence at their new URLs, the observed-transition share would be 30/425 (0.071) instead of 27/422 (0.064). The other three URL changes involve one transition whose new listing only adds a query parameter, one entry legacy-observed-only at both dates, and one entry era-unobservable at both dates.

\section{Interpretation and Reproducibility}

In November, 400 of 549 era-observable entries expose only legacy evidence. The Foundation's adoption account concerns early vendor support and explicitly notes backward compatibility~\cite{aaif2026ecosystemresponds}. Our sample measures the interfaces of registry-listed entries. For a client that needs broad coverage of this population, the observed legacy-observed-only share supports retaining initialization fallback, consistent with the specification's dual-era client behavior~\cite{mcp2026versioning}. A longitudinal QUIC study found outdated draft versions in the wild for up to four years and recommends that implementations keep parsing the identifiers they once shipped~\cite{Kempf_2026}. Because QUIC endpoints ignore unknown transport parameters, a stale parameter identifier there costs a feature; between a modern-only MCP client and a legacy-only server, the mismatch prevents the connection. The present result supports the protocol-level fallback decision; a separate client-interoperability study can measure its engineering cost and application-level reliability.

\paragraph{Coverage and authentication.}
The panel represents eligible remote entries in the October registry snapshot. It excludes source-only servers, noneligible transports, template URLs, and later registrations. Authentication blocks era observation for 292 entries in October and 275 in November. The estimand is therefore unauthenticated endpoint behavior; credentialed behavior forms a separate population. Registry listings also weight publishers with many entries more heavily than publishers with few entries.

\paragraph{Measurement validity.}
The classifier validates the response fields and recognized errors specified by the paired probe. The tool-name diagnostic measures the two declared-client conditions in Section~\ref{sec:design}. A companion end-to-end interoperability study can extend this design with standard modern and legacy clients, the protocol-specified HTTP fallback behavior, a negative control, and executable tool behavior. Each wave remains a short observation window from the same probing environment, so reachability is reported as an observation status rather than a permanent service classification.

\paragraph{Dependence and time.}
Domain clusters approximate common operation imperfectly: shared hosting can merge unrelated operators, and an operator can use multiple domains. Cluster adjustment quantifies dependence under that proxy, while registry coverage and authentication selection remain distinct sampling dimensions. The observed-transition estimate conditions on continued era observability, and the observed concentration identifies where a registrable-domain/publisher rerun is most informative. For 17 of the 27 transitions, October modern support was unobserved because discovery returned HTTP 401 (Section~\ref{sec:posthoc}). Endpoint moves recorded in the registry explain few of the entries that became era-unobservable (Section~\ref{sec:posthoc}), but an operator can also change the implementation behind an unchanged URL or move it without updating the registry. The two dates estimate a paired monthly change and do not identify a transition curve or its causes. A third wave around 2026-12-24 is planned to extend the panel.

\paragraph{Data, code, and reproducibility.}
The public sampling source is the Official MCP Registry~\cite{mcpregistry}. The registry snapshot used for sampling, the sampled panel of 1,000 entries, timestamped per-entry observations from both waves, the registry version histories used in Section~\ref{sec:posthoc}, the probe and analysis scripts with the hashes recorded before each wave, and a SHA-256 manifest are available at \url{https://github.com/zwang-devv/mcp-stateless-migration} (release \texttt{v1.0}); the analysis scripts regenerate the reported counts and intervals from these files. Registry metadata is in the public domain under the registry's terms. The released observations keep only the name, title, and version from each server's self-description, record HTTP status codes and content types, and contain no other headers, session identifiers, or credentials. The text withholds domain names when discussing individual operators.

\section{Conclusion}

In a fixed panel of registry-listed remote endpoints, 0.729 of entries era-observable without authentication remain legacy-observed-only in November, with a descriptive domain-cluster interval of \CI{0.592}{0.824}. Among 422 initially legacy-observed-only entries era-observable at both dates, 27 show an observed transition to dual-era evidence, a share of 0.064 with descriptive domain-cluster interval \CI{0.016}{0.164}. The transition signal is concentrated in one domain: excluding it gives 11/406 (0.027). For 17 of the 27, October modern support was unobserved because discovery returned HTTP 401; the conservative baseline-observable sensitivity gives 10/386 (0.026). The clustered interval for the fixed-panel increase in modern-era evidence includes zero. An earlier measurement of the same registry four weeks after the revision also found legacy-only behavior dominant~\cite{sunkavalli2026deploymentgap}. Together, the results show that unauthenticated protocol-era visibility, authentication policy, and deployment concentration must be separated when assessing stateless-revision uptake. Clients seeking coverage of these endpoints should retain legacy initialization fallback.

\small
\bibliographystyle{plainnat}
\bibliography{references}
\end{document}